\documentclass[aps,prx,reprint,preprintnumbers,superscriptaddress,nofootinbib,longbibliography,floatfix]{revtex4-2}
\pdfoutput=1
\usepackage{rotating}
\usepackage{array}
\usepackage{amsmath}
\usepackage{amssymb}
\usepackage[normalem]{ulem}
\usepackage{slashed}
\usepackage{booktabs}
\usepackage[pdftex,table]{xcolor}
\usepackage{units}
\usepackage{xfrac}
\usepackage{float}
\usepackage{mathtools}
\usepackage{empheq}
\usepackage[]{units}
\usepackage{multirow}
\usepackage{amssymb}
\usepackage{url}
\usepackage{comment}
\usepackage{physics}
\usepackage{color,soul}
\usepackage{bbm}
\usepackage[caption=false]{subfig}
\usepackage{ulem}
\usepackage{diagbox}

\usepackage{hyperref}
\hypersetup{
  colorlinks=true,
  citecolor=blue,
  linkcolor=blue,
  urlcolor=blue
}

\newcommand{\cf}{\textsc{CaloFlow}}
\newcommand{\cgan}{\textsc{CaloGAN}}

\begin{document}
\title{Anomaly detection with flow-based fast calorimeter simulators}

\author{Claudius Krause}
\email{Claudius.Krause@oeaw.ac.at}
\affiliation{Institut f\"ur Theoretische Physik, Universit\"at Heidelberg, Philosophenweg 12, 69120 Heidelberg, Germany}
\affiliation{Institute of High Energy Physics (HEPHY), Austrian Academy of Sciences (OeAW), Dominikanerbastei 16, A-1010 Vienna, Austria}

\author{Benjamin Nachman}
\email{bpnachman@lbl.gov}
\affiliation{Physics Division, Lawrence Berkeley National Laboratory, Berkeley, CA 94720, USA}
\affiliation{Berkeley Institute for Data Science, University of California, Berkeley, CA 94720, USA}

\author{Ian Pang}
\email{ian.pang@physics.rutgers.edu}
\affiliation{NHETC, Department of Physics and Astronomy, Rutgers University, Piscataway, NJ 08854, USA}

\author{David Shih}
\email{shih@physics.rutgers.edu}
\affiliation{NHETC, Department of Physics and Astronomy, Rutgers University, Piscataway, NJ 08854, USA}

\author{Yunhao Zhu}
\email{zhu.yunha@northeastern.edu}
\affiliation{NHETC, Department of Physics and Astronomy, Rutgers University, Piscataway, NJ 08854, USA}
\affiliation{Department of Physics, Northeastern University, Boston, MA 02115, USA}

\begin{abstract}
Recently, several normalizing flow-based deep generative models have been proposed to accelerate the simulation of calorimeter showers. Using \cf~as an example, we show that these models can simultaneously perform unsupervised anomaly detection with no additional training cost. As a demonstration, we consider electromagnetic showers initiated by one (background) or multiple (signal) photons.  The \cf~model is designed to generate single-photon showers, but it also provides access to the shower likelihood.  We use this likelihood as an anomaly score and study the showers tagged as being unlikely.  As expected, the tagger struggles when the signal photons are nearly collinear, but is otherwise effective.  This approach is complementary to a supervised classifier trained on only specific signal models using the same low-level calorimeter inputs.  While the supervised classifier is also highly effective at unseen signal models, the unsupervised method is more sensitive in certain regions and thus we expect that the ultimate performance will require a combination of these approaches.

\end{abstract}

\maketitle

\section{Introduction}
In 2012, the final piece of the Standard Model, the Higgs boson, was discovered at the Large Hadron Collider (LHC) by the ATLAS~\cite{ATLAS:2012yve} and CMS~\cite{CMS:2012qbp} collaborations. Despite this milestone, compelling theoretical and experimental reasons continue to drive the search for physics Beyond the Standard Model (BSM). Regrettably, the extensive search programs conducted by ATLAS~\cite{atlasexoticstwiki,atlassusytwiki,atlashdbspublictwiki}, CMS~\cite{cmsexoticstwiki,cmssusytwiki,cmsb2gtwiki}, and LHCb~\cite{lhcbtwiki} at the LHC have yet to yield conclusive evidence for new BSM physics. Given the impossibility of conducting dedicated searches for every conceivable BSM scenario, most LHC searches target specific signals derived from theoretical priors, leaving substantial portions of the LHC phase space unexplored. This limitation has spurred the development of more model-agnostic search strategies, with the hope of detecting BSM physics at the LHC. The advent of deep learning has given rise to various model-agnostic anomaly detection methods designed to explore uncharted territories of the LHC phase space --- for reviews and original references, see e.g.~\cite{hepmllivingreview, Kasieczka:2021xcg,Aarrestad:2021oeb,Karagiorgi:2021ngt,Feickert:2021ajf}.

Normalizing flows~\cite{tabak_flows,dinh2014nice,razende_flows,dinh2016density,flows_review,CNFs} represent a class of deep learning models particularly valuable for generative modeling and density estimation tasks. A normalizing flow is characterized by a parametric diffeomorphism $f_\theta$ mapping between a latent space, with a known distribution $\pi(z)$, and a data space of interest with an analytically unknown distribution $p(x)$. In the context of a conditional normalizing flow, this transformation becomes $f_\theta(x|c)$, where $c$ denotes the conditional inputs to the flow. It is defined through a series of invertible functions, parameterized by $\theta$, that can be trained by maximizing the log-likelihood of the data following the change of variables formula:
\[\log\bigl(p(x|c)\bigr) = \log(\pi(f_\theta(x|c))) + \log\Bigl|\det\left(\mathcal{J}\bigl(f_\theta(x|c)\bigr)\right)\Bigr|,\]
where $\mathcal{J}\bigl(f_\theta(x|c)\bigr)$ represents the Jacobian of the transformation $f_\theta(x|c)$. The allowable transformations must also have a computationally tractable Jacobian, ideally efficient to compute, and the probability density of the base distribution $\pi(z)$ must be known. A common choice for $\pi(z)$ is the standard normal distribution.


Recently, normalizing flows have found successful applications in fast calorimeter simulation tasks~\cite{Krause:2021ilc,Krause:2021wez,Krause:2022jna,Buckley:2023rez, Diefenbacher:2023vsw,Pang:2023wfx,Ernst:2023qvn}. Moreover, normalizing flows have demonstrated comparably excellent performance across various tasks within high energy physics~\cite{Nachman:2020lpy,Gao:2020vdv,Bothmann:2020ywa,Gao:2020zvv,Bellagente:2020piv,Stienen:2020gns,Bieringer:2020tnw,Bellagente:2021yyh, Hallin:2021wme,Bister:2021arb,Butter:2021csz, Winterhalder:2021ngy, Butter:2022lkf,Verheyen:2022tov, Leigh:2022lpn, Butter:2022vkj,Hallin:2022eoq,Heimel:2022wyj,Backes:2022vmn,Leigh:2022lpn,Raine:2023fko,Sengupta:2023xqy, Ackerschott:2023nax,Heimel:2023mvw,Heimel:2023ngj,Bierlich:2023zzd,Das:2023bcj}. In this paper, we demonstrate the utility of these flow-based generative models as unsupervised anomaly detectors for identifying BSM physics in calorimeter shower data. Specifically, we apply \cf\ \cite{Krause:2021ilc}, a flow-based fast calorimeter simulation model, to single-photon showers from a 
new sampling calorimeter version~\cite{krause_2023_10393540} of the \cgan~dataset~\cite{nachman2017electromagnetic,Paganini:2017hrr,Paganini:2017dwg}. In this context, single-photon showers serve as the background events, while the signal events consist of photon showers originating from a generic BSM particle $\chi$ that undergoes the decay $\chi \to \gamma\gamma$. The $\chi$ particle is taken to be invisible and interacts only indirectly with the calorimeter through its decay products. By training \cf~to maximize the log-likelihood when evaluated on background events, we are able to detect the signal events based on a cut on the log-likelihood.
We focus on achieving signal sensitivity, but the approach could be combined with a number of background estimation strategies~\cite{Nachman:2020lpy,Kasieczka:2021xcg}.

While we believe this is the first unification of simulation and anomaly detection, both subjects have been well-studied with machine learning.  Many deep generative models have been studied for calorimeter simulation~\cite{Paganini:2017hrr,Paganini:2017dwg,deOliveira:2017rwa,Erdmann:2018kuh,Erdmann:2018jxd,ATL-SOFT-PUB-2018-001,Belayneh:2019vyx,Vallecorsa:2019ked,SHiP:2019gcl,Chekalina:2018hxi,Carminati:2018khv,Vallecorsa:2018zco,Musella:2018rdi,Deja:2019vcv,ATLAS:2022jhk,ATLAS:2021pzo,ATLAS:2022jhk,Buhmann:2021lxj,Buhmann:2021caf,Krause:2021ilc,Krause:2021wez,Mikuni:2022xry,Buckley:2023rez,Krause:2022jna,Diefenbacher:2023vsw,Cresswell:2022tof,Hashemi:2023ruu,Liu:2023lnn,Diefenbacher:2023prl,Buhmann:2023bwk,Mikuni:2023tqg,Acosta:2023zik,Ernst:2023qvn}, including a number of proposals developed on the CaloChallenge datasets~\cite{calochallenge}, and they are also now being integrated into experimental workflows~\cite{ATLAS:2021pzo}.  We focus on normalizing flows since they give direct access to the likelihood.  This information can also be extracted from diffusion models~\cite{Mikuni:2023tok} and it would be interesting to compare approaches in the future.  For anomaly detection, unsupervised methods have been extensively studied (e.g. Refs.~\cite{Hajer:2018kqm,Farina:2018fyg,Heimel:2018mkt} and many others) and also include density-based approaches~\cite{Aarrestad:2021oeb}.  Like the density-based methods, we use the likelihood directly as the anomaly score.\footnote{
Note that this is not unique and is sensitive to how the data are represented/preprocessed~\cite{2020arXiv200608545K,Le_Lan_2021,Kasieczka:2022naq}.}
 
There may be other ways of reusing the generative model for BSM searches, including fine-tuning supervised models based on particular signal hypotheses.    

This paper is organized as follows. In Sec.~\ref{sec:data}, we describe the calorimeter setup and the datasets that were used during training/evaluation. In Sec.~\ref{sec:caloflow}, we explain how \cf~is used as an unsupervised anomaly detector by placing cuts on the log-likelihood of background and signal showers evaluated using \cf. In Sec.~\ref{sec:results}, we include the results of performing unsupervised anomaly detection with \cf. Finally, we conclude in Sec.~\ref{sec:conclusion}.

\section{Dataset}
\label{sec:data}

For this study, we decided to create a new, more realistic sampling calorimeter version~\cite{krause_2023_10393540} of the \cgan~dataset. The original dataset included energy contributions from both active and inactive calorimeter layers, whereas our new dataset includes only energy contributions from the active layers as would be available in practice. The sampling fraction of our new calorimeter setup is $\sim20\%$. The simple detector is a three-layer, liquid argon (LAr) sampling calorimeter cube with 480mm side length that is inspired by the ATLAS LAr electromagnetic calorimeter~\cite{CERN-LHCC-96-041}. It is simulated as follows: \textsc{Geant4}~\cite{Agostinelli:2002hh,1610988,ALLISON2016186} is used to generate particles and simulate their interaction with our calorimeter using the \textsc{Ftfp\_Bert} physics list based on the Fritiof~\cite{Ganhuyag:1997gz,Nilsson-Almqvist:1986ast,Andersson:1986gw,Andersson:1996xi} and Bertini intranuclear cascade~\cite{Guthrie:1968ue, Bertini:1971xb, Karmanov:1979if} models with the standard electromagnetic physics package~\cite{1462617}. While we use this new simulator to create a dataset for anomaly detection, we expect it should be more generally useful for a broad variety of tasks.

\begin{figure}
    \centering
    \includegraphics[width=\columnwidth]{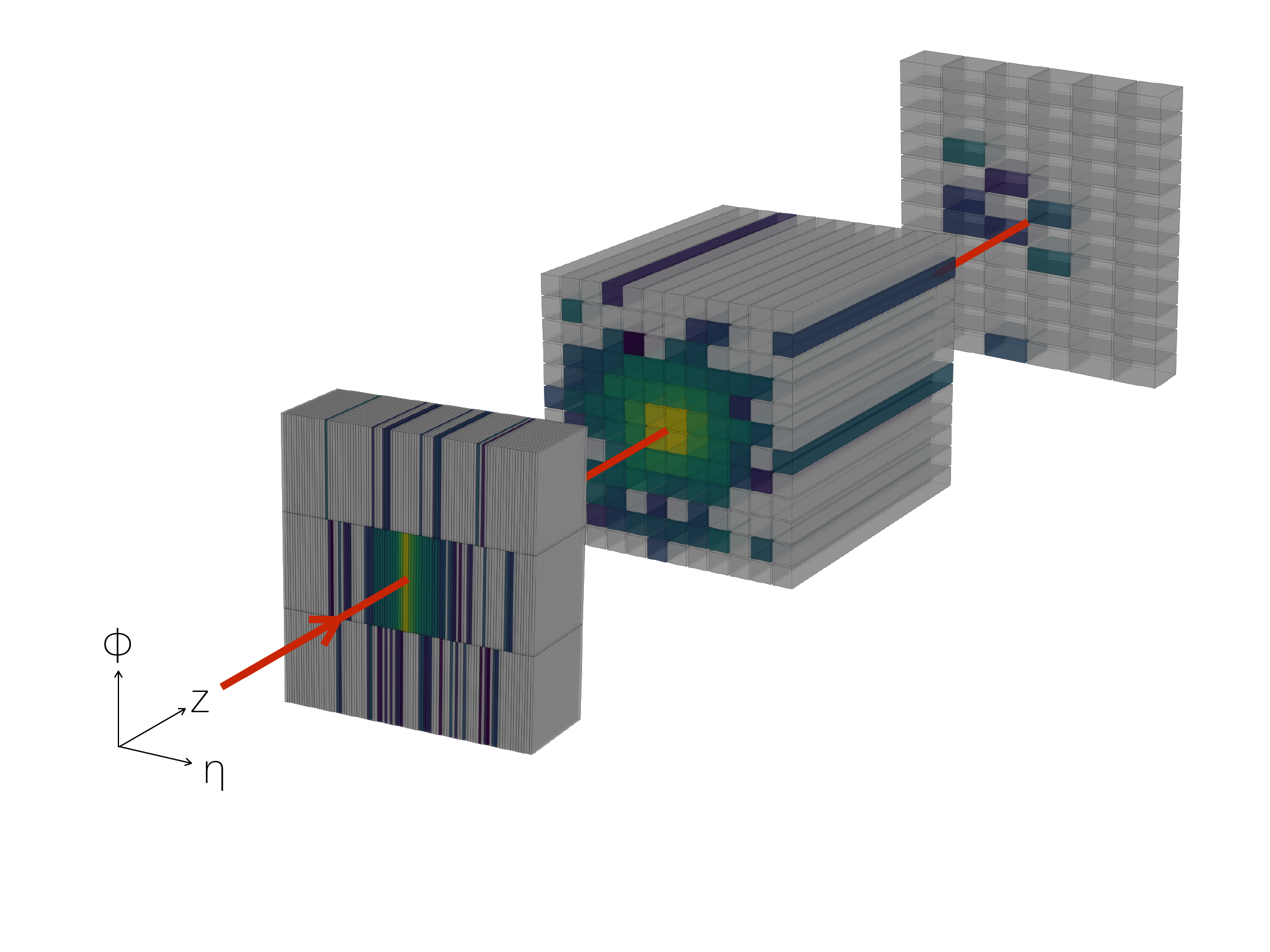}
    \caption{Three-dimensional representation of the shower in the \cgan~calorimeter, figure taken from ~\cite{Paganini:2017hrr,Paganini:2017dwg}. Not-to-scale separation among the longitudinal layers is
added for visualization purposes.}
    \label{fig:3d}
\end{figure}

The calorimeter showers are represented as three-dimensional images that are binned in position space. In this representation, the calorimeter shower geometry is made up of voxels (volumetric pixels) and the details of the calorimeter voxel dimensions are included in Table \ref{tab:calorimeter_specs}. Figure~\ref{fig:3d} (taken from~\cite{Paganini:2017hrr,Paganini:2017dwg})   shows the three-dimensional representation of a shower in the \cgan~calorimeter. The three longitudinal layers are separated in the figure for visualization purposes. In this work, the center of the detector is taken to be at $z=0$ m, while the front face of layer 0 is positioned at $z=1$ m which is consistent with an ATLAS-like configuration.

\begin{table}[!b]
\begin{center}
\begin{tabular}{|c|c|c|c|c|}
\hline
\begin{tabular}{c}Layer \\index\end{tabular} & \begin{tabular}{c}$z$ length \\ (mm) \end{tabular} &\begin{tabular}{c}$\eta$ length \\ (mm) \end{tabular} & \begin{tabular}{c}$\phi$ length \\ (mm) \end{tabular} & \begin{tabular}{c}Number \\of voxels\end{tabular}\\
\hline\hline
0 & 90 & 5 & 160 & $3\times96$ \\ \hline
1 & 347 & 40 & 40 & $12\times12$ \\ \hline
2 & 43 & 80 & 40 & $12\times6$ \\ \hline
\end{tabular} 
\caption{Dimensions of a calorimeter voxel. The positive $z-$axis (radial direction in full detector) is the direction of particle propagation, the $\eta$ direction is along the proton beam axis, and $\phi$ is perpendicular to $z$ and $\eta$. For the number of voxels, the first (second) number is the number of bins in the $\phi$ ($\eta$) direction (e.g., $12\times 6$
refers to 12 $\phi$ bins and 6 $\eta$ bins).}
\label{tab:calorimeter_specs}
\end{center}
\end{table}

For the background dataset (single-photon calorimeter showers), we generate 100k showers with incident energies $E_\text{inc}$ uniformly distributed in the range \([1,100]\) GeV. This dataset was used as the training dataset for \cf~where we used a train/validation split of 70\%/30\%. A second independent dataset of 100k showers with the same range of $E_\text{inc}$ was generated and used for evaluation.

To obtain the signal datasets, we defined the hypothetical $\chi$ particle in \textsc{Geant4} which has the same properties as the $\eta^0$ particle except having a different mass and being invisible to the detector. Next, we generate multiple sets of 100k showers that originate from $\chi\to\gamma\gamma$ decay at 10 chosen fixed displacements from the center of detector along the $z-$axis. The chosen fixed displacements are $z_i\in\{0.33, 0.66, 1.00,   1.04, 1.08, 1.16, 1.24, 1.32, 1.40, 1.44\}$ m. Note that the first two displacements are located in front of the calorimeter, while the last eight displacements are located at distinct positions within the calorimeter. The energy of the $\chi$ particle was fixed at 50 GeV. Such a hypothetical scenario might arise from the decay of a 100 GeV particle (close to e.g. the Higgs boson mass), which is at rest, to a pair of $\chi$ particles.

To study how the mass $m_\chi$ affects the anomaly detection performance, we generated signal datasets with different $m_\chi \in \{5\times10^{-3},\ 5\times10^{-2},\ 5\times10^{-1},\ 5\}$ GeV. For each choice of $m_\chi$, we generate 100k showers at each of the 10 fixed displacements. The results for these fixed displacement signal datasets are shown in Section~\ref{sec:fixed_displacement}.

To consider particles with fixed lifetimes, we construct new signal datasets using the fixed displacement samples. In particular, we use the probability that the $\chi$ particle decays at various positions along the $z-$axis to determine the proportion of showers originating from decays at each of the 10 fixed displacements. Generating directly with fixed lifetimes 
would have been too inefficient, since 
we are discarding decays after the calorimeter volume, which can happen often for these lifetimes that we consider.
More details of the fixed lifetime signal datasets are included in Section~\ref{sec: vary_lifetime}.

We also generated 100k signal showers with kinetic energy\footnote{We distinguish between kinetic energy and incident energy for massive particles. The incident energy is the sum of the kinetic energy and rest mass energy of the particle. This distinction only becomes noticeable in the case of $m_\chi = 5$ GeV.} uniformly distributed in the range [1,100] GeV for a $\chi$ particle with $m_\chi = 5\times10^{-3}$ GeV and lifetime $\tau = 1.00$ ns. A second set of 100k signal showers was generated with the same range of kinetic energies, but for a $\chi$ particle with $m_\chi = 5$ GeV and lifetime $\tau = 1.00$ ns. Each of these two datasets (together with the background dataset) was used to train a supervised classifier described in Section~\ref{sec:compare_supervised}. Using a range of kinetic energies ensured that we obtain showers with a range of $E_\text{inc}$ such that the $E_\text{inc}$ would not be used as a discriminating factor by the supervised classifier. This allows for a fairer comparison between performance of the supervised classifier and our unsupervised method since \cf~learns the likelihood conditioned on $E_\text{inc}$.

\section{\cf}
\label{sec:caloflow}
\cf~\cite{Krause:2021ilc,Krause:2021wez} is an approach to fast calorimeter simulation based on conditional normalizing flows. In the context of fast calorimeter simulation, \cf~uses a two-flow method to learn to generate the voxel level shower energies $\vec{\mathcal{I}}$ conditioned on the corresponding incident energies of the showers $E_\text{inc}$ denoted by $p(\vec{\mathcal{I}}|E_\text{inc})$. Flow-I is constructed to learn the probability density of calorimeter layer energies\footnote{The layer energy of a given calorimeter layer is the sum of all the voxel energies in that layer.} $E_i$ conditioned on the incident energy $p_1(E_0, E_1, E_2|E_\text{inc})$, while Flow-II is designed to learn the probability density of the normalized voxel level shower energies conditioned on the calorimeter layer energies and incident energies $p_2\left(\hat{\mathcal{I}}| E_0, E_1, E_2, E_\text{inc}\right)$. By normalized, we mean that the voxel energies in each layer are made to sum to unity. The dimensions of the conditional inputs and outputs of the two flows are shown in Table~\ref{tab:conditional}. Importantly, these flows were trained only using single photon showers.

One important difference from the original \cf~is that in this application to anomaly detection we do not have direct empirical access to $E_\text{inc}$. For a given shower, we do not know a priori what the corresponding $E_\text{inc}$ is and would instead have to use a reconstructed estimate of $E_\text{inc}$ which we denote as $E^{(\text{rec})}_\text{inc}$. In this work, we use a simple regression method to reconstruct $E_\text{inc}$ given the total deposited energy in the calorimeter $E_\text{dep}$. The reconstructed incident energy is defined as $E^{(\text{rec})}_\text{inc} = \lambda E_\text{dep}$, where $\lambda$ is the mean of $E_\text{inc}/E_\text{dep}$ in the single photon training dataset. Note that the true $E_\text{inc}$ is still used for training \cf, while $E^{(\text{rec})}_\text{inc}$ is used when performing anomaly detection. 

The architecture and training of \cf~are outlined in Appendix~\ref{sec:arch_training}. Some modifications were made to the original \cf, while most of the main algorithm remains unchanged. 

\begin{table}
    \centering
    \begin{tabular}{|c|c|c|c|c|} \hline
         & Conditionals &\begin{tabular}{@{}c@{}}Dim of \\ conditional\end{tabular} & Output &\begin{tabular}{@{}c@{}}Dim of \\ output\end{tabular}  \\ \hline\hline
Flow-I & $E_{\rm inc}$ & 1  & $E_0, E_1, E_2$ & 3\\ \hline 
Flow-II & $E_0, E_1, E_2, E_{\rm inc}$  & 4 & $\hat{\mathcal{I}}$ & 504\\ \hline
    \end{tabular}
    \caption{The conditional inputs for each flow, and the features whose probability distributions are the output of each flow.}
    \label{tab:conditional}
\end{table}

\subsection{Unsupervised anomaly detection with \cf}
After training \cf~on background single-photon showers, we evaluate the log-likelihood \(\log p\left(\hat{\mathcal{I}}, E_i|E^{(\text{rec})}_\text{inc}\right)\) of the background and signal showers by using the trained flows. Using both Flow-I and Flow-II, we are able to obtain the full log-likelihood: 

{\small
\begin{align}
\log p\left(\hat{\mathcal{I}}, E_i|E^{(\text{rec})}_\text{inc}\right) =
\underbrace{\log p_1(E_i|E^{(\text{rec})}_\text{inc})}_{\text{Flow-I}} + \underbrace{\log p_2\left(\hat{\mathcal{I}}| E_i, E^{(\text{rec})}_\text{inc}\right)}_{\text{Flow-II}}\,.
\end{align}
}
\begin{figure}
    \centering
    \includegraphics[width=0.9\columnwidth]{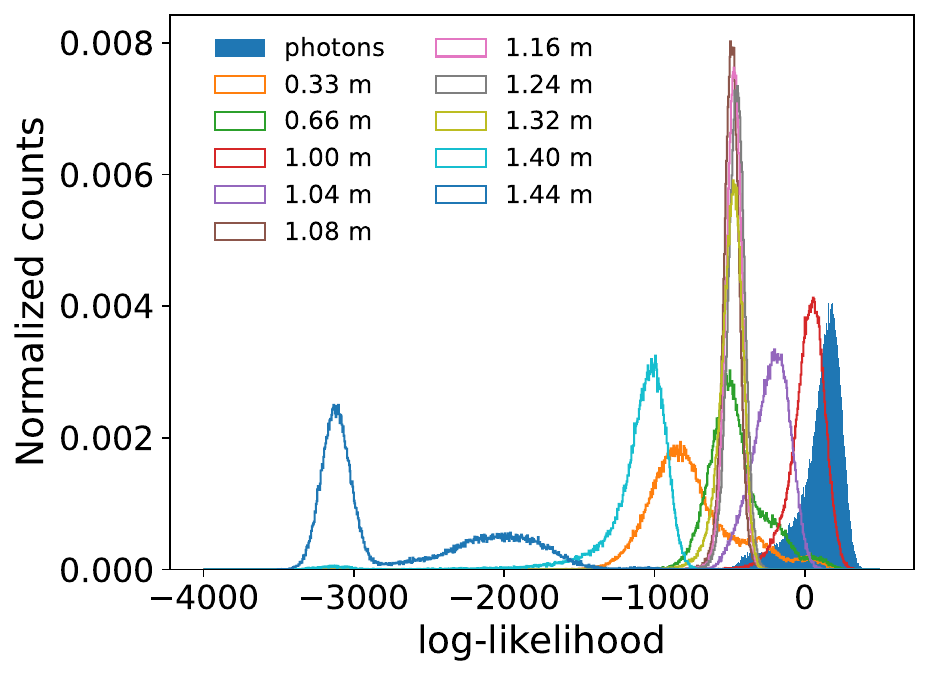}
    \caption{Plot of log-likelihood comparison between signal showers from $\chi$ decays at the 10 fixed displacements (lines) and background photon showers (filled). This example is for a $\chi$ particle with $m_\chi = 5$ GeV.}
    \label{fig:cf_ad}
\end{figure}

Figure~\ref{fig:cf_ad} shows an example of a plot of the full log-likelihood of the signal showers from $\chi$ decays at the 10 fixed displacements and background photon showers. 

We see from Figure~\ref{fig:cf_ad} that the signal showers generally have log-likelihoods that are distinguishable from that of the background showers. Hence, we can use \cf~as an unsupervised anomaly detector by placing cuts on the full log-likelihood to discriminate signal from background showers. The results are detailed in Section~\ref{sec:results}. Though not explained here, the main features found in Figure~\ref{fig:cf_ad} can be understood from the discussion in Section~\ref{sec:fixed_displacement}.

Despite the possibility of data-MC differences affecting the sensitivity of our approach, we note that the accuracy would not be affected since one would presumably apply standard downstream background estimation. Furthermore, experiments usually calibrate their fast simulation and those calibrations could be applied to improve the sensitivity.
\label{sec:method}

\section{Results}
\label{sec:results}
\subsection{Decay at fixed displacement}
\label{sec:fixed_displacement}
In this section, we study the effect of varying the displacement from the center of the detector at which the decay occurs on the evaluated likelihood. In reality, particles do not decay at fixed positions, but instead the probability of a particle decaying at a given displacement from where it was created is related to its lifetime $\tau$. Nevertheless, studying the showers produced by the particle $\chi$ at fixed decay positions is interesting from an experimental perspective, and doing so also helps us interpret the more physical results in Section \ref{sec: vary_lifetime}.

\begin{figure}
    \centering
    \includegraphics[width=\columnwidth]{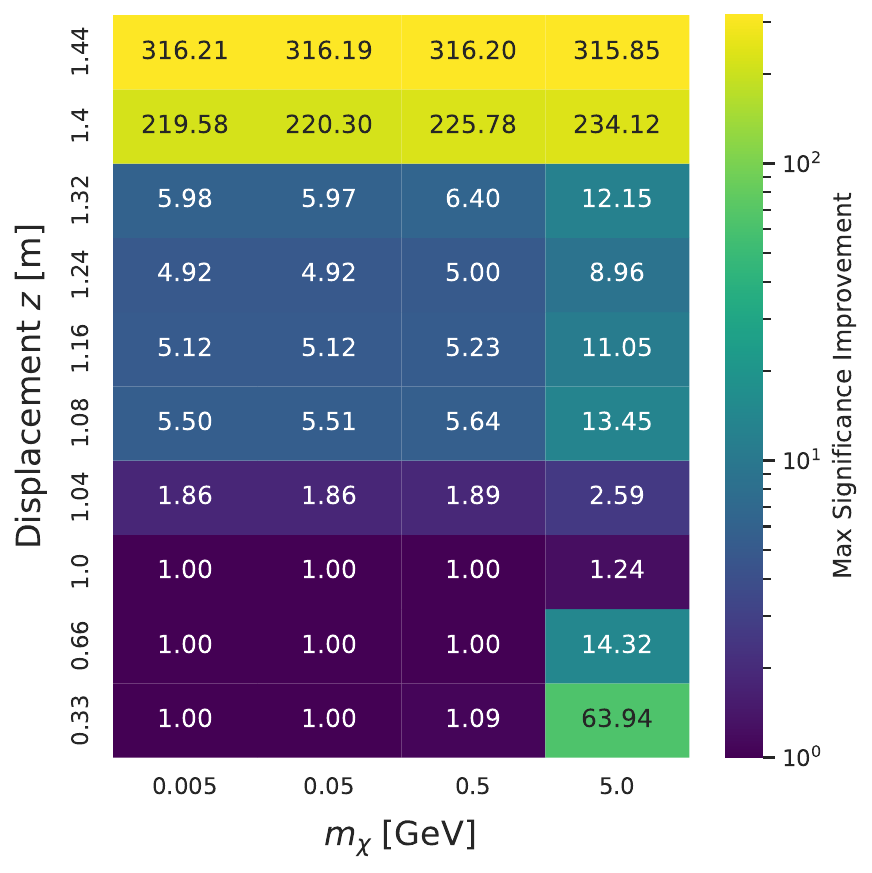}
    \caption{Heatmap of maximum significance improvement $\left(\text{tpr}/\sqrt{\text{fpr}}\right)$ for different masses of the $\chi$ particle $m_\chi$ and the 10 fixed displacements where the decay occurs.}
    \label{fig:fixed_disp_heatmap}
\end{figure}

\begin{figure*}[ht]
\includegraphics[width=0.5\columnwidth]{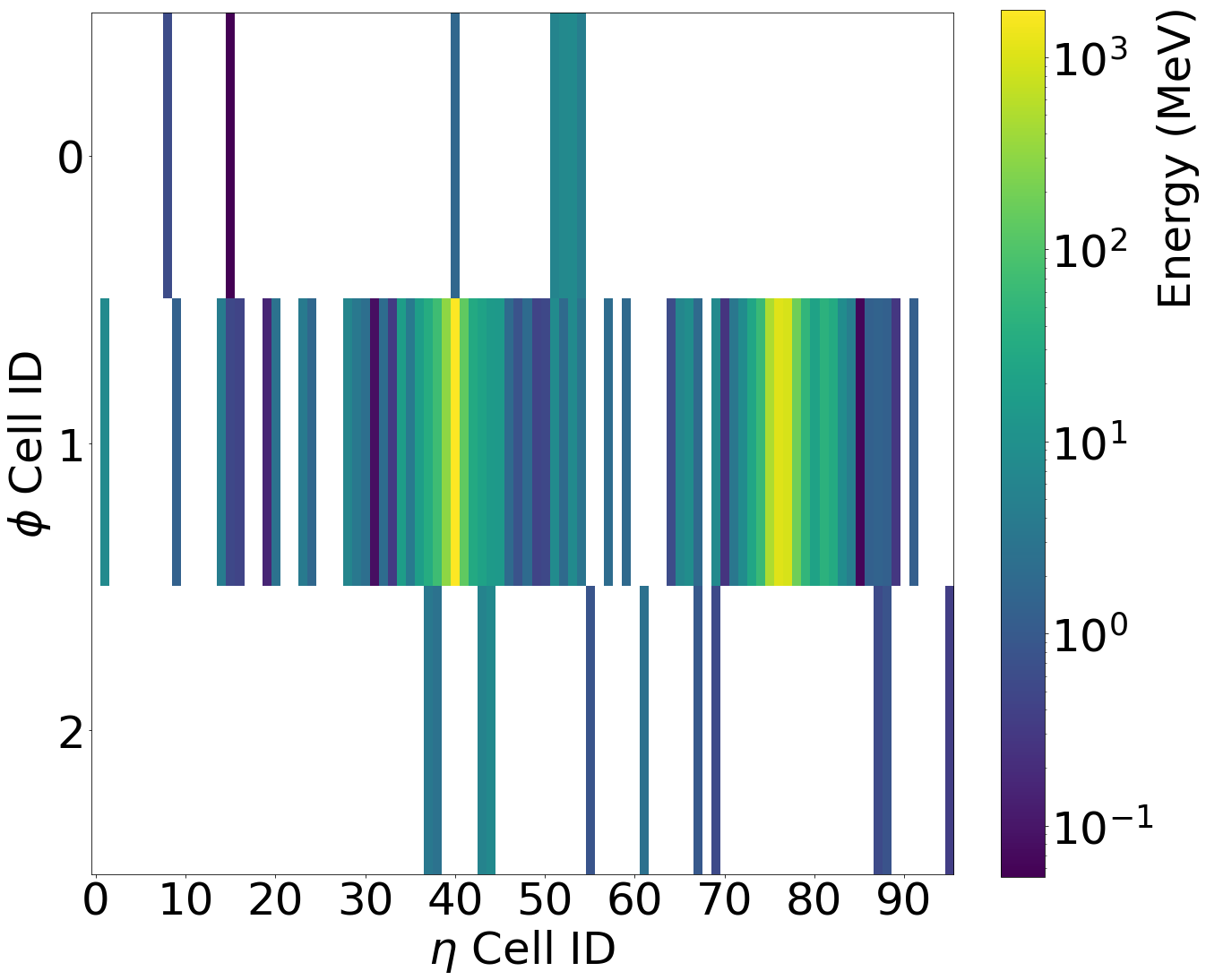} \includegraphics[width=0.5\columnwidth]{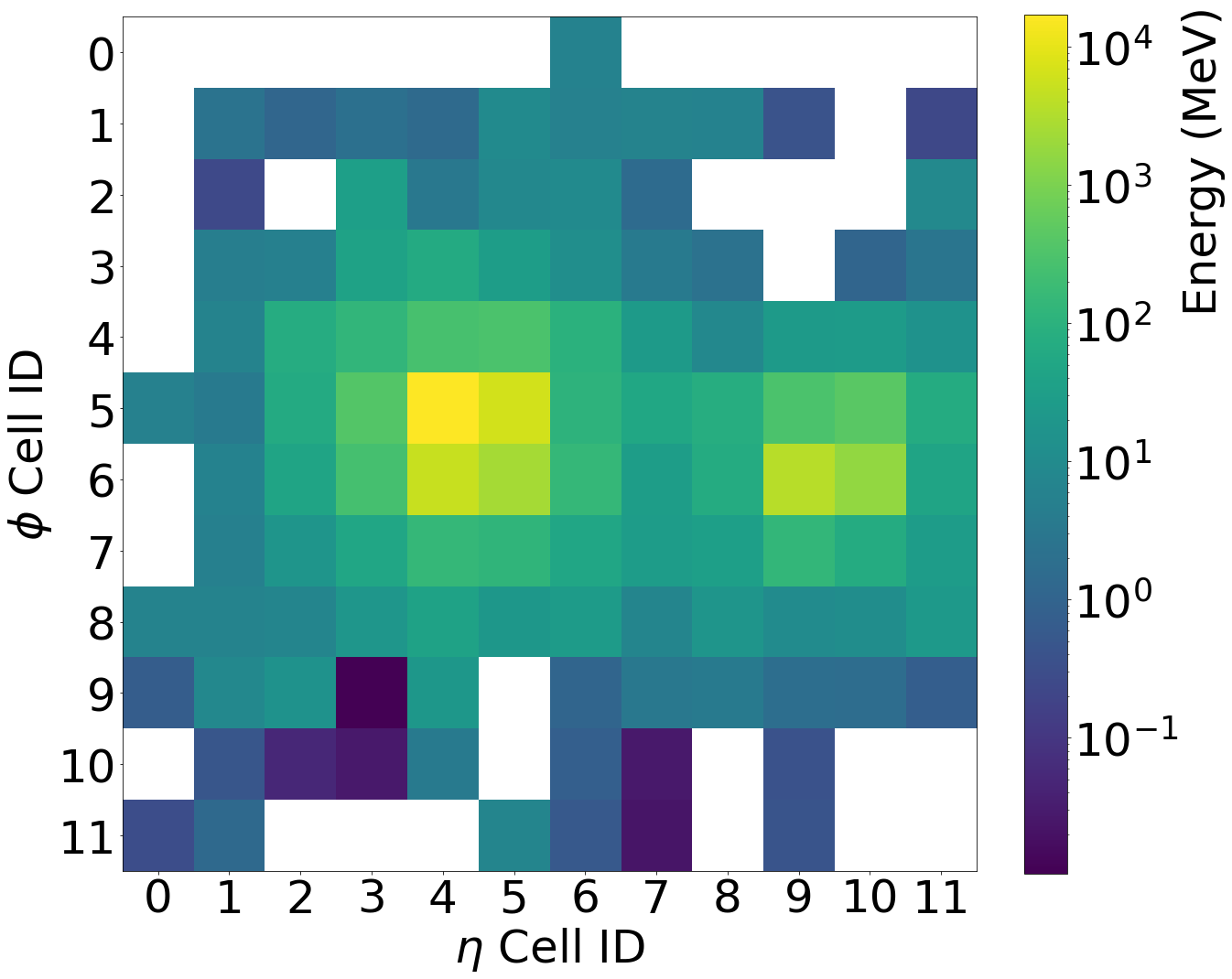}\includegraphics[width=0.5\columnwidth]{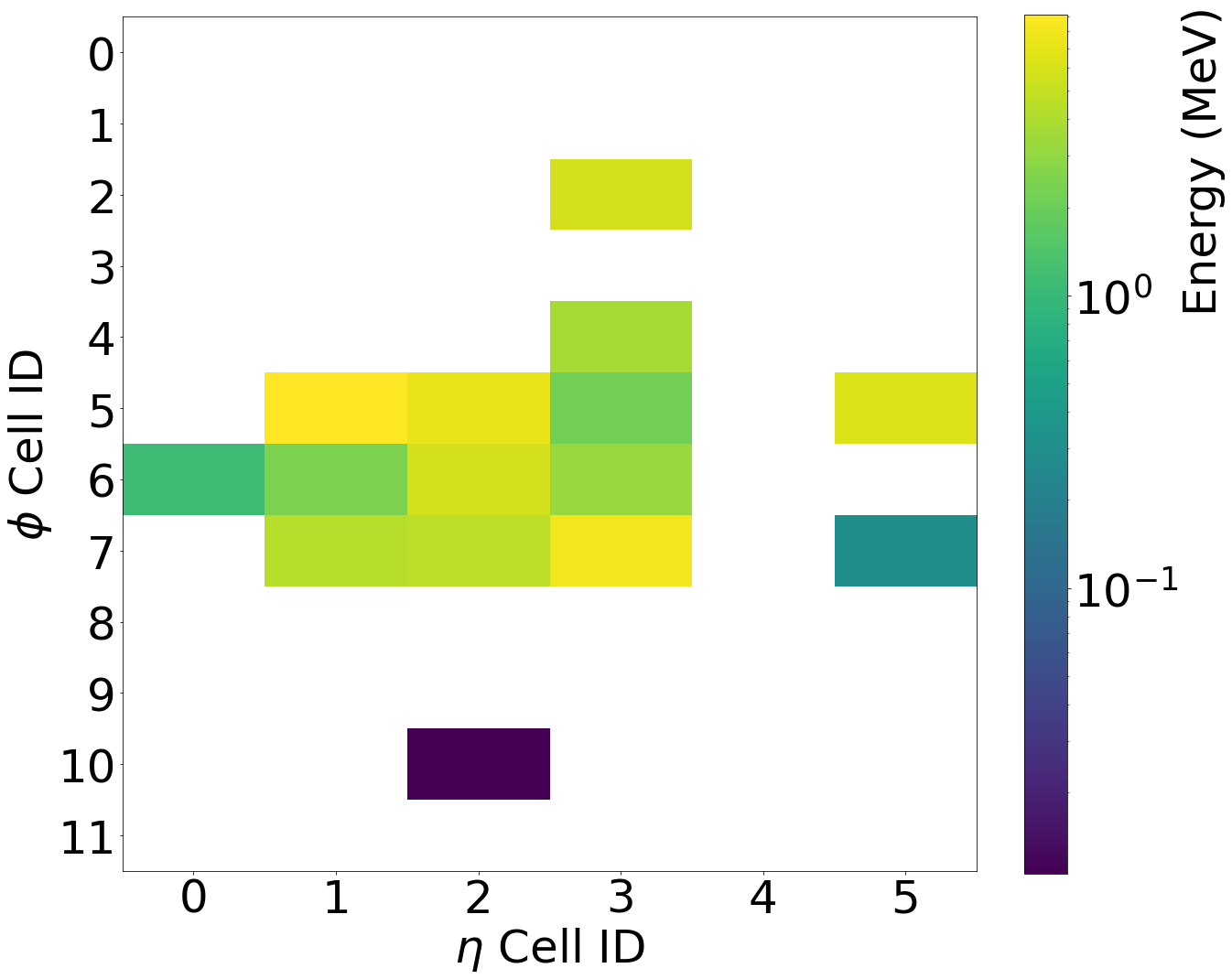}\\
\includegraphics[width=0.5\columnwidth]{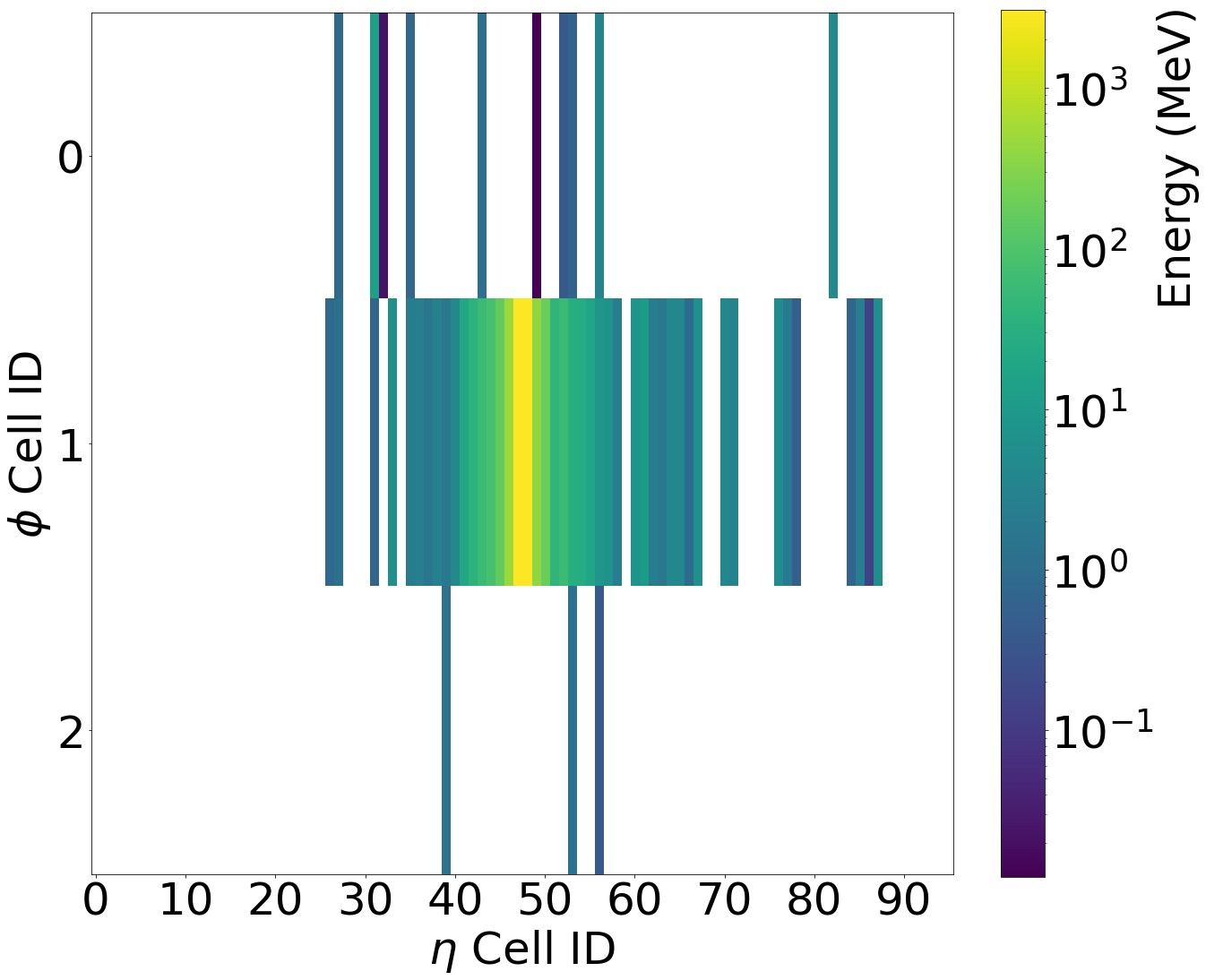}\includegraphics[width=0.5\columnwidth]{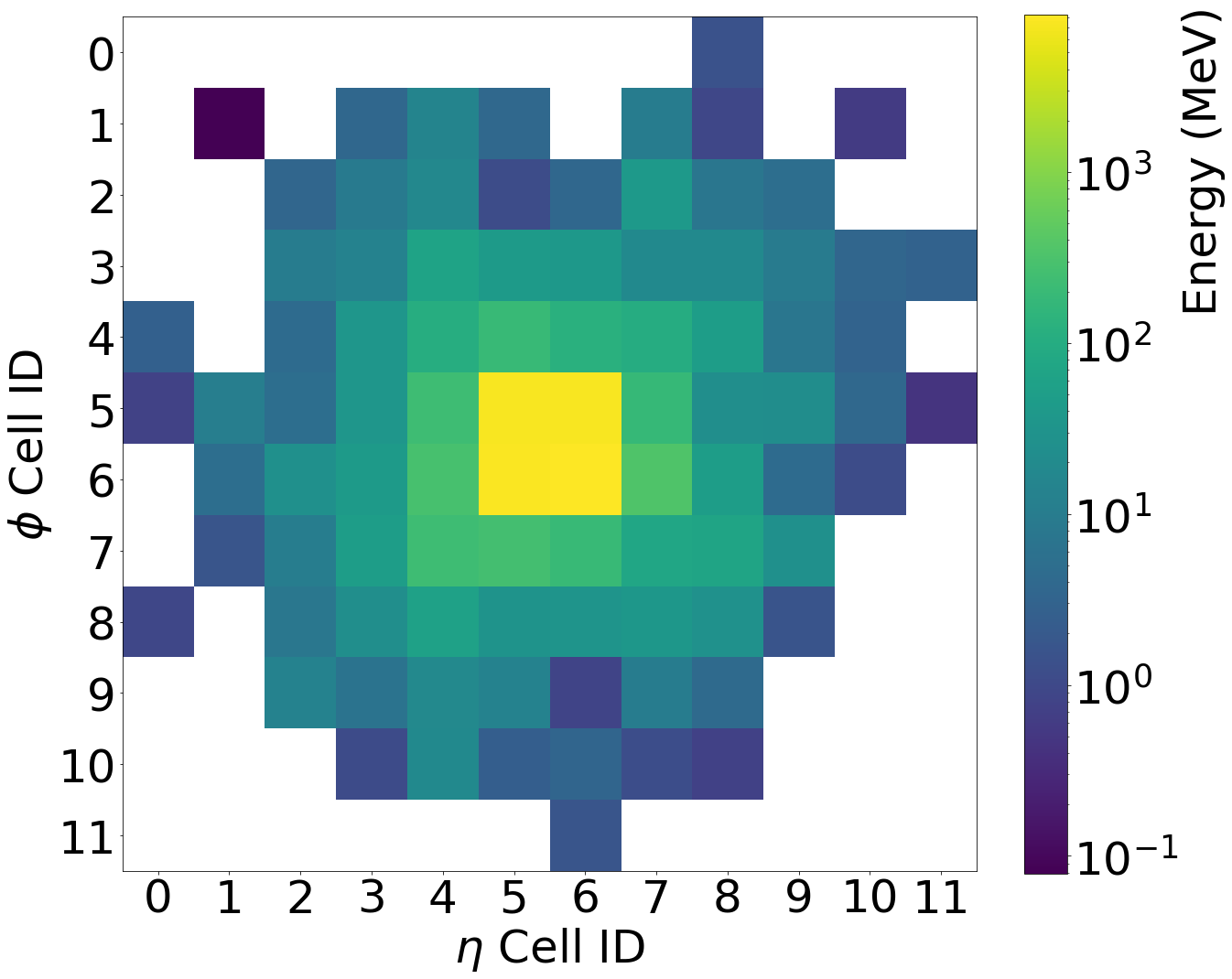}\includegraphics[width=0.5\columnwidth]{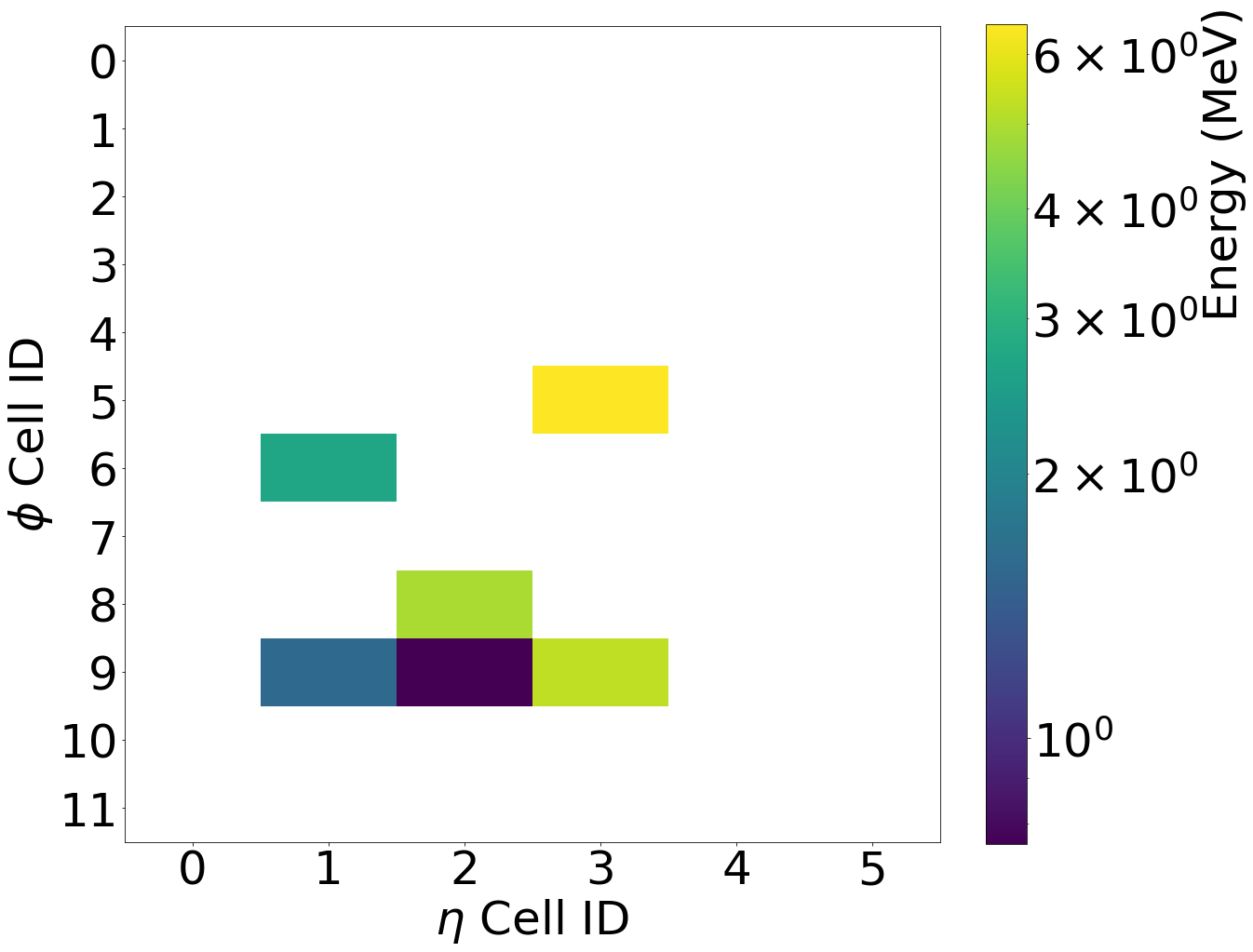}

\caption{Top row: Example of a shower with two distinct blobs of energy originating from a 5 GeV $\chi$ particle that decayed at $z<1.0$ m. Bottom row: Example of a typical photon shower with centralized energy core. The energy deposition in each of the three layers is shown here with layer 0 on the left, layer 1 in the center, and layer 2 on the right.}
\label{fig:blobs}
\end{figure*}

Our performance metric for anomaly detection will be the significance improvement which is defined as the signal efficiency (i.e.~true positive rate or tpr) divided by the square root of the background efficiency (i.e.~false positive rate or fpr). The maximum significance improvement corresponds to the best possible\footnote{This is signal model dependent, but still provides a useful bound on the achievable performance.} multiplicative factor by which the signal significance can be improved with a cut on the anomaly score.  Figure~\ref{fig:fixed_disp_heatmap} shows a heatmap of maximum significance improvement for each of the four different particle masses $m_\chi$, and the 10 fixed displacements where the decay occurs. In general, we find that showers from decays at larger $z$ are more anomalous. However, there is a clear exception in the case of $m_\chi = 5$ GeV, where the showers originating from decays that occur before the $\chi$ particle reaches the calorimeter (e.g.,~$z =$ 0.33 m, 0.66 m) are more anomalous than those from decays occurring at the front face of the calorimeter (i.e.,~$z=1.0$ m). This is due to the fact that the 5 GeV particle is less boosted compared to the other lighter particles that we consider in this study. As a result, the decay of the 5 GeV particle often results in a wider angle between the produced pair of photons which \cf~is better able to distinguish from the background single-photon showers. See Figure~\ref{fig:blobs} for an example of such a shower and comparison with a regular photon shower. On the other hand, if a decay occurs right at the front face of the calorimeter, there is insufficient time for the pair of photons to propagate and create two distinct blobs of energy in the calorimeter. The sudden jump in maximum significance improvement for all $m_\chi$ when
going from $z = 1.32$ m to $z = 1.4$ m is due to the discretization of the calorimeter voxel geometry in the longitudinal direction described in Section~\ref{sec:data}. The transition between the second and third longitudinal layers occurs at $z=1.437$ m. Hence, showers from decays at $z = 1.4$ m, which have more energy deposited in the third layer, are flagged as significantly more anomalous.

For the three lowest $m_\chi$, we find that \cf~struggles to detect signal showers that originate from decays occurring before the front face of the calorimeter (i.e.,~$z\leq 1.0$ m). These lower mass particles are highly boosted which results in showers from highly collimated photon pairs that are more similar to the background photon showers.

\subsection{Varying lifetime}
\label{sec: vary_lifetime}

To assess the performance of our unsupervised anomaly detector on realistic scenarios, as mentioned in Section~\ref{sec:data}, we construct new datasets consisting of showers from $\chi$ particles with fixed rest frame lifetimes. In particular, for each chosen mass $m_\chi$ and lifetime $\tau$, we have a total of 100k events that are made up of showers from $\chi$ decays at fixed displacements. The proportion of showers associated to decays at each fixed displacement is determined based on the lifetime $\tau$. Particles that decay after the calorimeter volume are not included in the events since they are not detected within the calorimeter.

The probability for a particle to survive for time $t$ before decaying is given by $P_s(t) = \exp(-\frac{t}{\gamma \tau})$, where the relativistic boost factor $\gamma = E_\text{particle}/M_\text{particle}$. Equivalently, the probability that a particle decays before reaching displacement $z$ is given by $P_d(z) = 1-\exp(-\frac{z}{c\tau\sqrt{\gamma^2-1}})$. Since we consider 10 fixed displacements indexed by $i \in \{1,2,...,10\}$, the fraction of the total showers that originates from $\chi$ decay at displacement $z_i$ is set to be $\hat w_i$ which is defined by

\[w_i  = \left\{
    \begin{array}{ll}
        P_d(z_1),\quad i=1 \\
        P_d(z_i)-P_d(z_{i-1}),\quad \text{otherwise}
    \end{array}\right.\]
    \[\text{and}\quad\hat w_i = \frac{w_i}{\sum_{j=1}^{10}w_j}\]

\begin{figure}[h]
    \centering
    \includegraphics[width=0.5\textwidth]{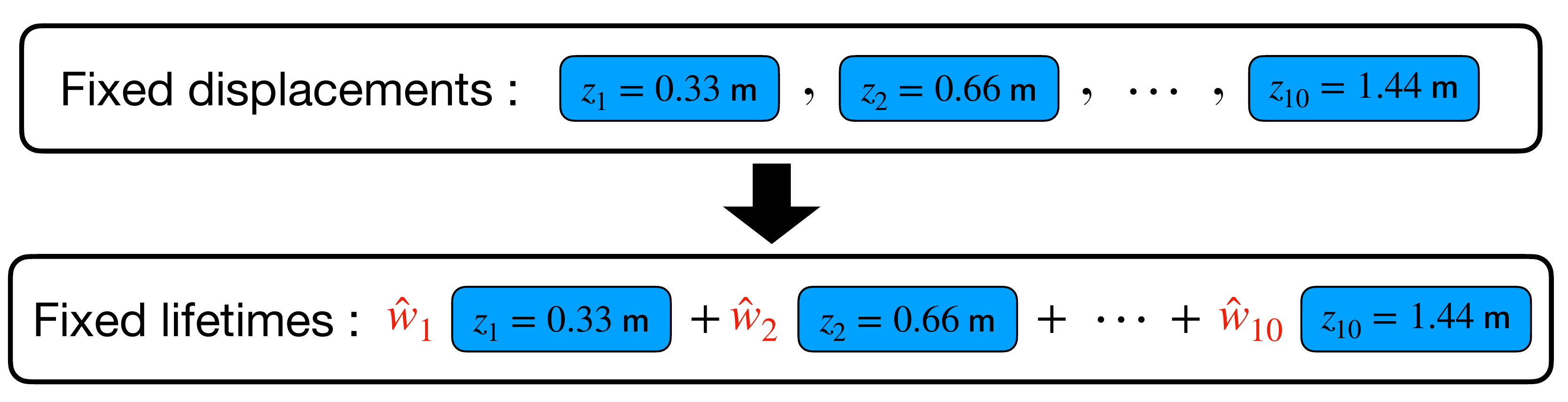}
    \caption{Visualization of how the fixed lifetime datasets are constructed by sampling from the fixed displacement datasets. The fraction of total showers in each fixed lifetime dataset originating from $\chi$ decay at displacement $z_i$ is set to be $\hat{w}_i$.}
    \label{fig:gen_construct}
\end{figure}

In other words, the number of showers originating from the decay at $z = z_i$ is equal to $10^5 \times \hat w_i$\footnote{In some cases, we had to round $\hat w_i$ to ensure that the total number of showers is equal to 100k.}. In this work, we consider three possible lifetimes $\tau \in \{0.01, 0.1, 1\}$ ns for each choice of $m_\chi$. A visualization of how the fixed lifetime datasets are constructed from showers in the fixed displacement datasets is shown in Figure~\ref{fig:gen_construct}. In Table~\ref{tab:percent_flight_length}, we show the percentage of $\chi$ decays that occur before/within the calorimeter volume and also the average flight length of the $\chi$ particle for different $m_\chi$ and lifetimes $\tau$.

\renewcommand{\arraystretch}{1.5}
\begin{table*}[h!]
\centering
\begin{tabular}{|c||c|c|c|c|}
\hline
\backslashbox{$\tau$}{$m_\chi$} & $5 \times 10^{-3}$ GeV & $5 \times 10^{-2}$ GeV & $5 \times 10^{-1}$ GeV& $5$ GeV\\ 
\hline \hline
0.01 ns& 5\% / 30 m & 38\% / 3 m & 99\% / 0.3 m & 100\% / 0.03 m \\ 
\hline
0.1 ns& 0.5\% / 300 m & 5\% / 30 m & 38\% / 3 m & 99\% / 0.3 m \\ 
\hline
1  ns& 0.05\% / 3000 m & 0.5\% / 300 m & 5\% / 30 m & 38\% / 3 m \\ 
\hline
\end{tabular}
\caption{Percentage of $\chi$ decays (left number) that occur before/within the calorimeter volume, and the average flight length (right number) of $\chi$ particles for different $m_\chi$ and lifetimes $\tau$.}
\label{tab:percent_flight_length}
\end{table*}
\renewcommand{\arraystretch}{1}

\begin{figure*}[h]
\includegraphics[width=0.7\columnwidth]{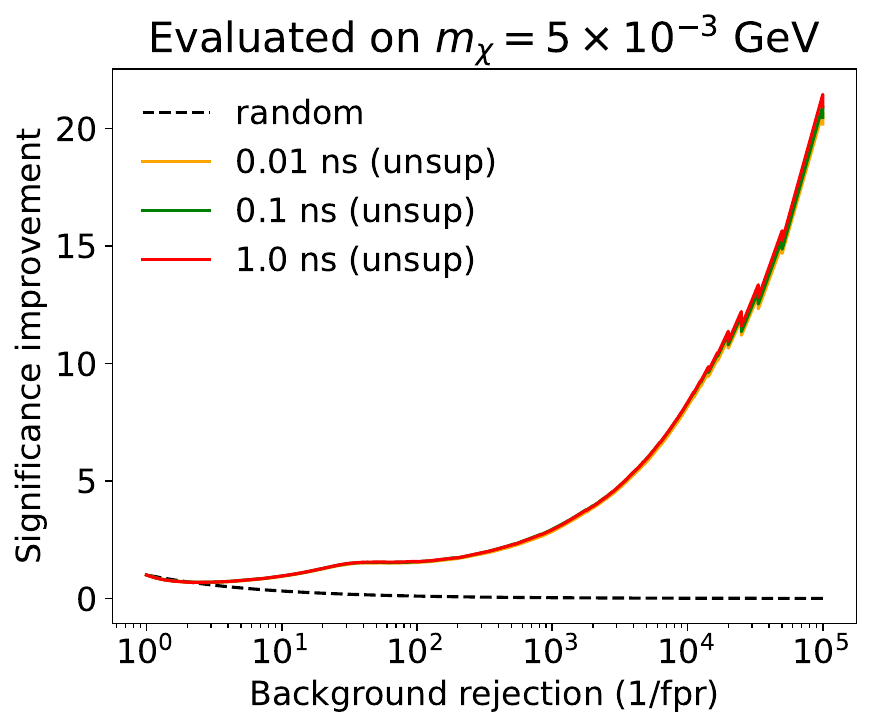}\hspace{0.1\textwidth} \includegraphics[width=0.7\columnwidth]{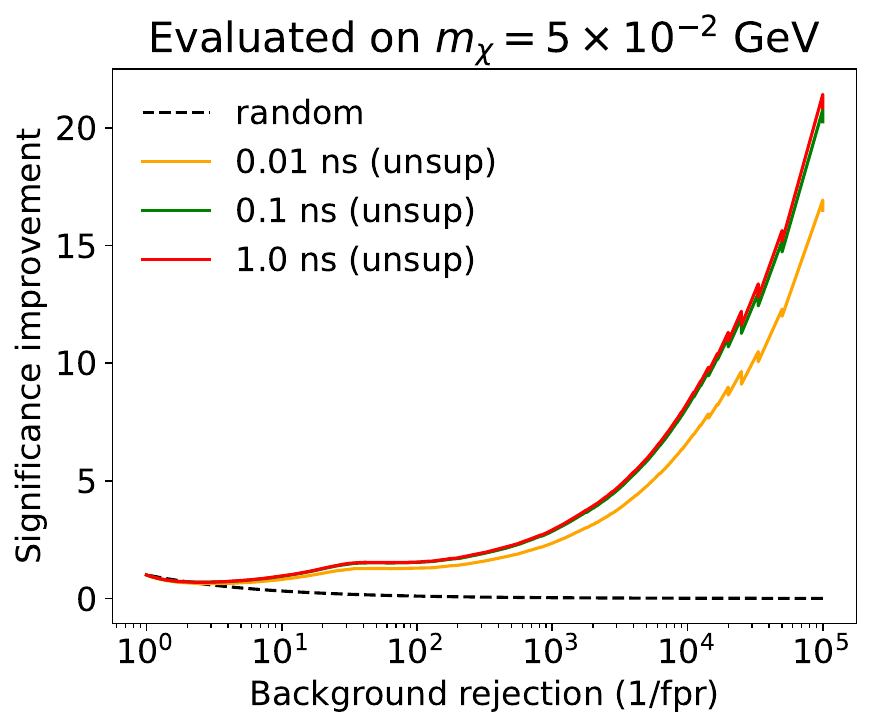}\\
\includegraphics[width=0.7\columnwidth]{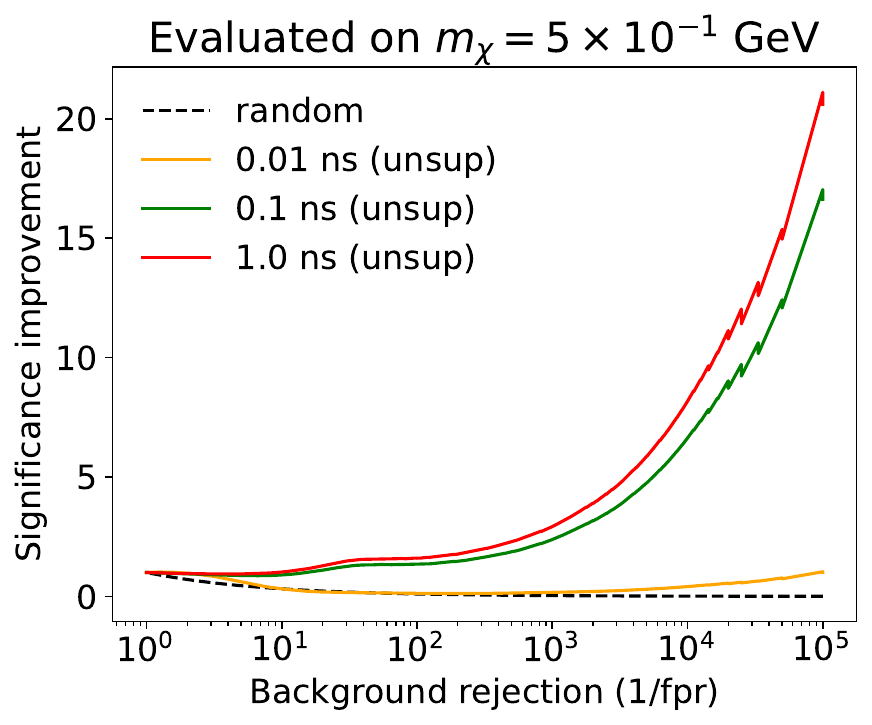}\hspace{0.1\textwidth}\includegraphics[width=0.7\columnwidth]{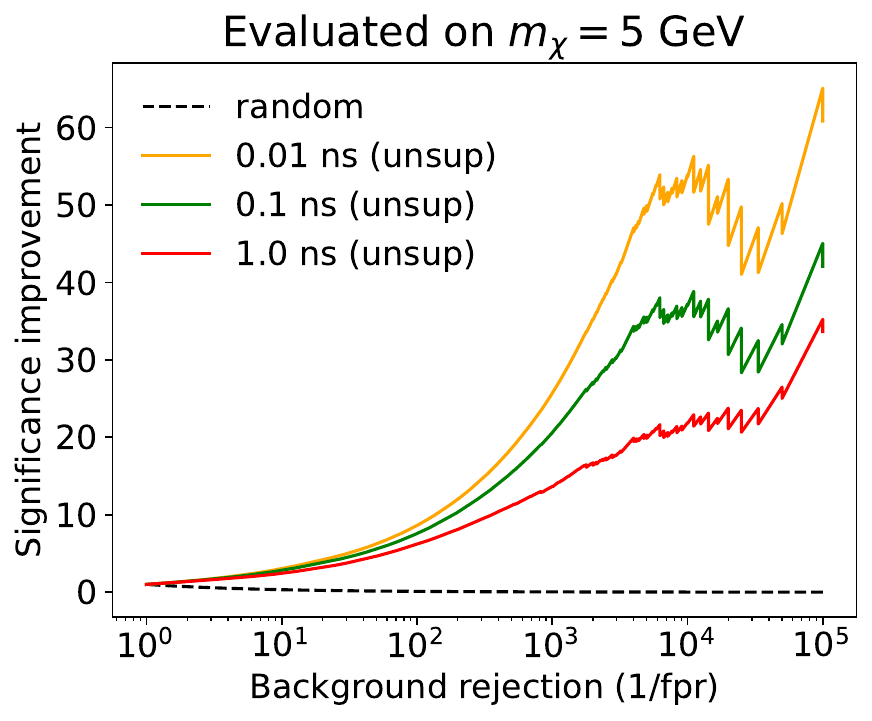}

\caption{Plots of significance improvement of our unsupervised (unsup) \cf~anomaly detector as a function of background rejection (1/false positive rate) for various types of signal showers originating from $\chi$ particles with different $m_\chi$ and $\tau$. Top left: $m_\chi = 5\times10^{-3}$ GeV, top right: $m_\chi = 0.05$ GeV, bottom left: $m_\chi = 0.5$ GeV, bottom right: $m_\chi = 5$ GeV. The performance of a random classifier is drawn with black dashed lines to serve as a baseline.}
\label{fig:vary_lifetime}
\end{figure*}

Figure~\ref{fig:vary_lifetime} shows the significance improvement of our \cf~anomaly detector as a function of the background rejection (1/false positive rate) for various types of signal showers originating from $\chi$ particles with different $m_\chi$ and $\tau$. 

Overall, we find that \cf~is able to achieve significantly better performance compared to a random classifier which we take to be the baseline. One exception is for $(m_\chi, \tau) = (0.5 \ \text{GeV}, 0.01 \ \text{ns})$ where the performance closely matches that of the random baseline. The poorer performance here is due to the lower boost factor $\gamma$ for the $0.5$ GeV particle compared to the two lower particles masses which results in the majority of the decays occurring close to the center of the detector ($z=0$ m). Meanwhile, the mass is still low enough that the photons are not widely separated by the time they reach the calorimeter. As we have seen in Figure~\ref{fig:fixed_disp_heatmap}, such showers are mostly not detected as anomalous by \cf. At longer lifetimes, the showers are more anomalous as a larger proportion of them originate from decays occurring within the calorimeter. 

For the three lowest $m_\chi$, the significance improvement generally increases with background rejection. Also, we find that the significance improvement curves are similar across lifetimes for the two lowest $m_\chi$. In these cases, $\frac{z}{c\tau\sqrt{\gamma^2-1}}$ is small which implies that $P_d(z) \approx \frac{z}{c\tau\sqrt{\gamma^2-1}}$. Hence, the lifetime $\tau$ cancels out when computing $\hat w_i$. In other words, at fixed $m_\chi$, the large boost of these particles results in a similar proportion of particles decaying at a given fixed displacement before/within the calorimeter for different lifetimes $\tau$.

The best performance among all the (fixed mass and lifetime) signal models we considered in this study was achieved in the cases with the largest mass of $m_\chi = 5$ GeV. As explained in Section~\ref{sec:fixed_displacement}, showers from early decay of these more massive particles are quite anomalous according to \cf~due to the wider angle between the produced pair of
photons. In this case, going to higher lifetimes actually makes these showers slightly {\it less} anomalous (which is opposite to the trend seen at lower masses) since it gives the photons less time to separate before interacting with the detector. 

\begin{figure}
    \centering
    \includegraphics[width=\columnwidth]{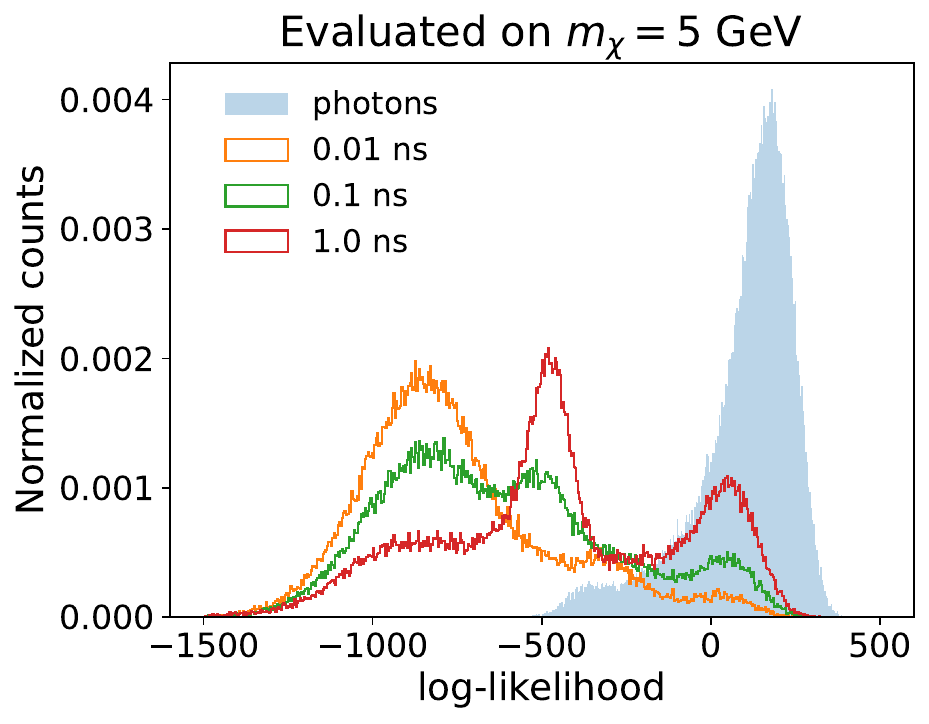}
    \caption{Plot of log-likelihood of $\chi$ showers for $m_\chi = 5$ GeV at the different $\chi$ particle lifetimes.}
    \label{fig:5GeV_LL}
\end{figure}

There is a local maximum in each of the significance improvement curves for $m_\chi = 5$ GeV. To understand the local maximum, we have to look at the log-likelihood plot for $m_\chi = 5$ GeV shown in Figure~\ref{fig:5GeV_LL}. The peaks at the log-likelihoods of -1000 and -500 are due to showers from decays occurring at $z =$ 0.33 m and 0.66 m respectively which can also be seen in Figure~\ref{fig:cf_ad}. Sliding the cut on log-likelihood in the direction of decreasing log-likelihood is equivalent to increasing the background rejection (1/fpr). Notice that sliding the cut in the direction of decreasing log-likelihood across peaks in the $\chi$ shower log-likelihood curve would create a local maximum in the plot of significance improvement vs background rejection (1/fpr) since the tpr decreases faster than the increase in $1/\sqrt{\rm fpr}$.

\subsection{Comparison with supervised anomaly detection}
\label{sec:compare_supervised}

The previous section showed that the unsupervised anomaly detector has broad coverage across the various model parameters.  An important question to ask is how this compares to a dedicated search for the $\chi\rightarrow\gamma\gamma$ signal.  For a particular signal model, we would expect the dedicated search to outperform the unsupervised approach.  However, it is not possible to have a dedicated search for every possible signal and so the key question to ask is how well a supervised model trained on one signal would perform on other signals not seen during training.

In this section, we compare the performance of our method against two supervised classifiers. Each supervised classifier was trained on a combined dataset with 100k signal showers and 100k background showers. The showers originate from particles with kinetic energy uniformly distributed in the range [1,100] GeV. The signal showers originate from $\chi$ particles with lifetime $\tau = 1.00$ ns. The first (second) supervised classifier was trained on a dataset with $m_\chi = 5\times10^{-3} \ (5)$ GeV. These models were chosen because they are sufficiently different that they would likely be covered by different dedicated searches.  It is thus interesting to ask if a search optimized for one of the models would still be sensitive to the other, since the unsupervised approach has some sensitivity to both models.

The supervised classifier is a fully-connected neural network with two hidden layer with 512 nodes each. We have a 505-dimensional input consisting of the voxel energies normalized by the reconstructed incident energy $\vec{\mathcal{I}}/E^{(\text{rec})}_\text{inc}$ (504-dim) and the reconstructed incident energy $E^{(\text{rec})}_\text{inc}$ (1-dim). The output layer returns a single number which is passed through a sigmoid activation function. All other activation functions are ReLUs. The supervised classifier was trained for a total of 50 epochs with a train/test/validation split of 60\%/20\%/20\%.  These parameters were not extensively optimized, but we found little gain from small variations in the setup. \cf~has a total of $\sim 38$M parameters, while the supervised classifier only has $\sim 522$k parameters. However, we found that increasing the number of parameters for the supervised classifier does not help to increase its performance. \cf~has a larger number of parameters because it has a more difficult task of learning the likelihood of the data.

\begin{figure*}[h]
    \centering
    \begin{minipage}{\textwidth}
        \centering
        \textbf{Top row: Supervised classifier trained on $\mathbf{m_\chi = 5\times 10^{-3}}$~GeV}
        
        \vspace{0.2cm}
        \includegraphics[width=0.4\textwidth]{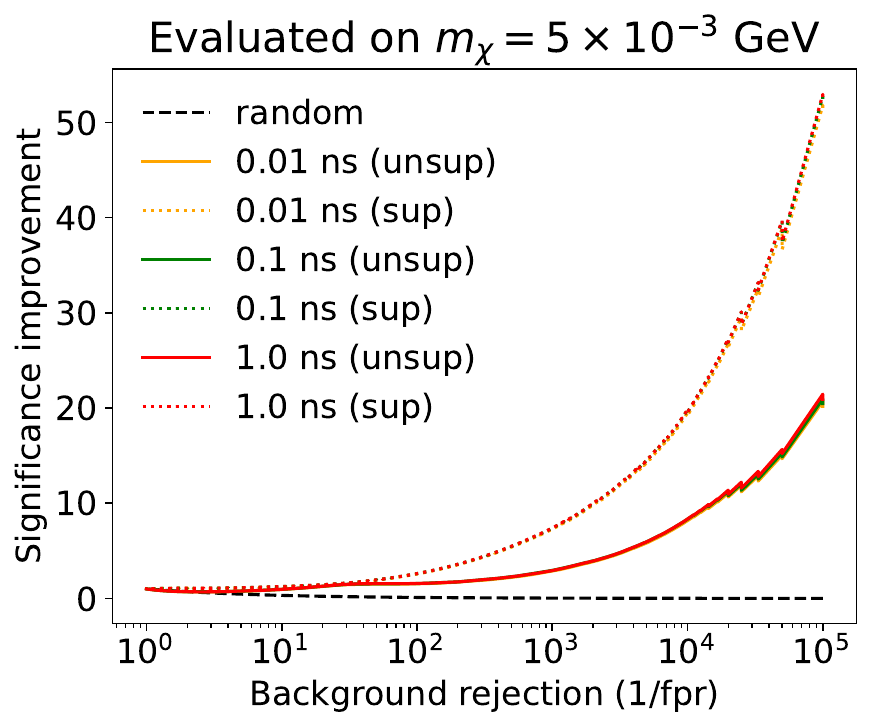}
        \hspace{0.05\textwidth}
        \includegraphics[width=0.4\textwidth]{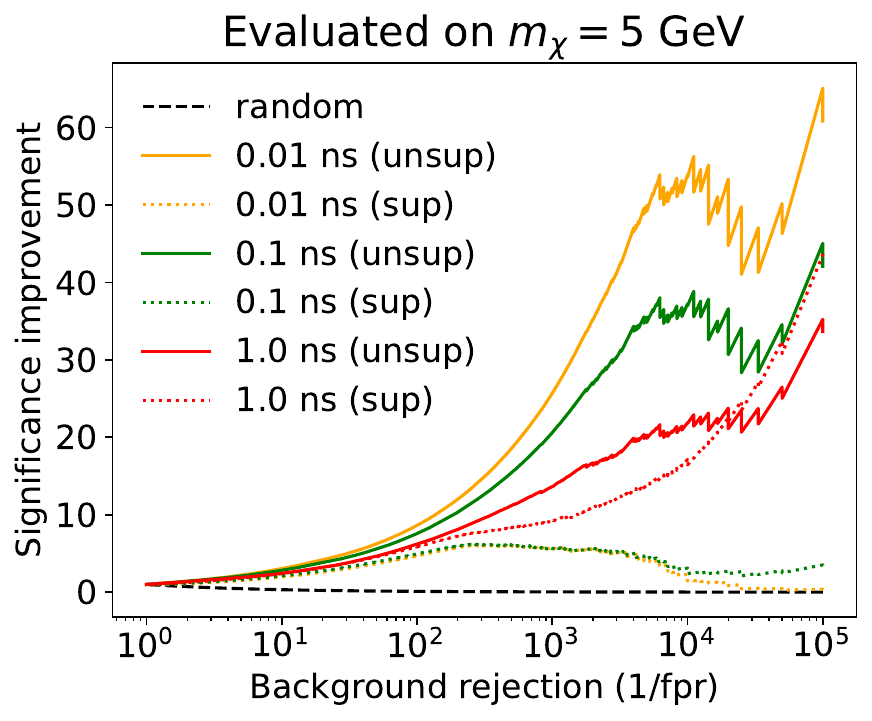}
    \end{minipage}
    
    \vspace{0.5cm}
    
    \begin{minipage}{\textwidth}
        \centering
        \textbf{Bottom row: Supervised classifier trained on $\mathbf{m_\chi = 5}$~GeV}
        
        \vspace{0.2cm}
        
        \includegraphics[width=0.4\textwidth]{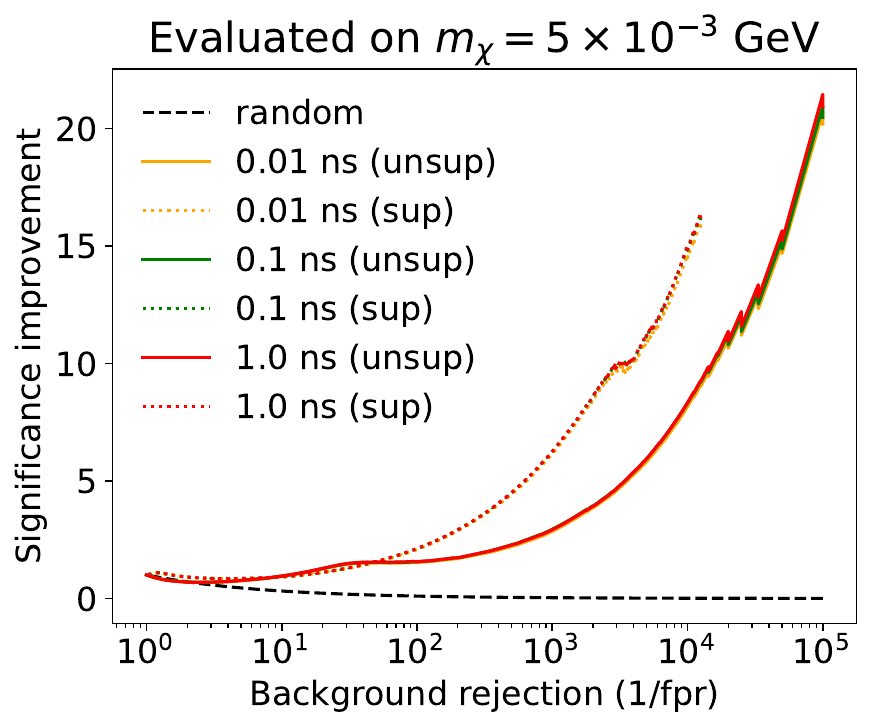}
        \hspace{0.05\textwidth}
        \includegraphics[width=0.4\textwidth]{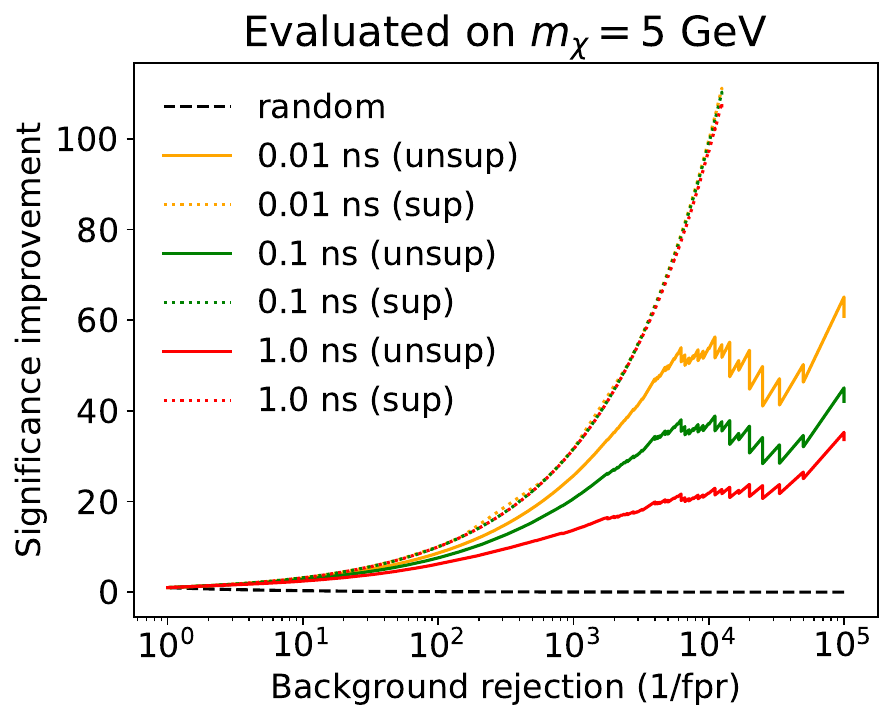}
    \end{minipage}
    
    \caption{Comparison plots of significance improvement of our \cf~anomaly detector as a function of background rejection (1/false positive rate). The plots for the performance for the $m_\chi = 5\times 10^{-3}$ GeV and  $m_\chi = 5$ GeV cases are shown on the left and right respectively. \textbf{Top row}: supervised classifier was trained on signal showers originating from $\chi$ particles with $m_\chi = 5\times10^{-3}$ GeV and lifetime $\tau = 1.00$ ns. \textbf{Bottom row}: supervised classifier was trained on signal showers originating from $\chi$ particles with $m_\chi = 5$ GeV and lifetime $\tau = 1.00$ ns. The performance of our unsupervised \cf~anomaly detector (unsup) is shown as solid lines, while that of the supervised classifier (sup) is shown as dotted lines. The performance of a random classifier is drawn with black dashed lines to serve as a baseline.}
    \label{fig:supervised-comparison}
\end{figure*}



After training the supervised classifiers, they were evaluated on the signal showers discussed in Section~\ref{sec: vary_lifetime}. Figure~\ref{fig:supervised-comparison} shows the comparison plots between the performance of our \cf~anomaly detector (unsup) and the supervised classifiers (sup). 
We see from the figure that the supervised classifier always outperforms \cf~on showers originating from $\chi$  particles that have the same
mass as what it was trained on (see top left and bottom right plots). However, the performance of the supervised classifier usually decreases when evaluated on signal showers originating from $\chi$  particles that have very different
masses from what it was trained on (see top right and bottom left plots). Whether the supervised classifier or \cf~achieves better performance depends on $m_\chi$. 

\begin{itemize}

\item For the upper right plot (trained on $m_\chi=5\times 10^{-3}$~GeV and evaluated on $m_\chi=5$~GeV), \cf~easily outperforms the supervised classifier as the supervised classifier did not see training examples of anomalous showers that are characteristic of larger $m_\chi$ (e.g.~early decay resulting in two blobs) and is unable to generalize its performance. 

\item However, in the lower left plot (trained on $m_\chi=5$~GeV and evaluated on $m_\chi=5\times 10^{-3}$~GeV) \cf~does not outperform the supervised classifier.  This is likely due to \cf~not being able to fully discriminate against signal showers that only decay after the first longitudinal layer (i.e.,~$E_0=0$ GeV), whereas such showers are usually perfectly distinguished by the supervised classifier.\footnote{Here, the point is that \cf~is not able to distinguish showers from late decays as well as the supervised classifier can. Nevertheless, for $m_\chi = 5\times 10^{-3}$ GeV, \cf~still has a better anomaly detection performance on late decay (larger $z$) showers compared to early decay (smaller $z$) showers.} As noise is added to the voxel energies when using \cf, this artificially causes $E_0 > 0$ GeV\footnote{We checked that a different treatment of the layers with zero energy deposition does not improve the performance.} and prevents such showers from being perfectly distinguished by \cf.
\end{itemize}


Finally, let us also comment on some other interesting features in Fig.~\ref{fig:supervised-comparison}.
When the classifier or anomaly detector is evaluated on $m_\chi=5\times 10^{-3}$~GeV (left column in Fig.~\ref{fig:supervised-comparison}), the performance is the same for all the lifetimes considered. This is because, for $m_\chi=5 \times 10^{-3}$ GeV, the particles are highly boosted, so  there is a similar proportion of the total number particles decaying at a given fixed displacement before/within the calorimeter for different lifetimes. (Keep in mind that we only consider particles decaying before/within the calorimeter.) Interestingly, the fully supervised classifier is also the same for all lifetimes for the bottom right plot (trained and evaluated on $m_\chi=5$~GeV). Here, the reason is that the signal showers look extremely different from the background photon showers and are perfectly distinguished by the classifier. Thus, we see that the significance improvement is equal to $1/\sqrt{\rm fpr}$.



This comparison of our unsupervised anomaly detection method against a supervised classifier highlights the potential limitation of model-specific anomaly detection as the supervised model is unable to generalize its excellent performance to signal that is too different from what it was trained on. We note that it is possible to train a supervised classifier on all the types of signal showers we have considered here. Doing so would likely result in the supervised classifier outperforming \cf~when evaluated on all the signal showers. However, the point is that even if one is to train a supervised classifier on a large number of signal types, it is impossible to exhaust the space of all possible signals. Hence, there may be an advantage in using model-agnostic, unsupervised anomaly detection methods such as the one we proposed in this work. This is especially true when using flow-based fast calorimeter simulators as no additional training has to be performed to use them as unsupervised anomaly detectors. 

\section{Conclusion and Outlook}
\label{sec:conclusion}


Using \cf~as an example, we demonstrated how fast calorimeter surrogate models with access to the data likelihood can double up as unsupervised anomaly detectors. 

By studying the anomaly detection performance of \cf~on showers from $\chi$ particle decays occurring at fixed displacements in the detector, we found that \cf~is generally more sensitive to signal showers from decays that occur deeper in the calorimeter. However, in the case of more massive, less highly boosted $\chi$ particles, \cf~still has significant discriminative power for showers from decays occurring in front of the calorimeter.

By reweighting the proportion of showers originating from decays at each fixed displacement, we constructed signal datasets corresponding to fixed particle lifetimes. We found that \cf~has discriminative power for most of the models we tested. In particular, \cf~achieves the best performance in the case with $m_\chi = 5$ GeV and $\tau = 0.01$ ns where the particle is less highly boosted and the majority of the particles decay close to the center of the detector.

Finally, we compared the performance of our unsupervised \cf~anomaly detector against a supervised classifier. We found that a supervised classifier trained on signal showers from highly boosted $\chi$ particles performed significantly poorer on showers from more massive, less highly boosted particles compared to our unsupervised method. When trained on signal showers from more massive $\chi$ particles and
applied to signal showers from less massive $\chi$ particles, the supervised classifier still performs
well.
This highlights the complementarity of different approaches and reaffirms the need to have a diversity of methods in order to achieve broad sensitivity.


\section*{Data and code availability}
The datasets used in this study can be found at~\cite{krause_2023_10393540} and the software to generate these datasets are located at \href{https://github.com/hep-lbdl/CaloGAN/tree/samplingEM}{https://github.com/hep-lbdl/CaloGAN/tree/samplingEM}. The machine learning software is at \href{https://github.com/Ian-Pang/AD\_with\_CF}{https://github.com/Ian-Pang/AD\_with\_CF}.

\section*{Acknowledgements}
CK would like to thank the Baden-W\"urttemberg-Stiftung for financing through the program \textsl{Internationale Spitzenforschung}, pro\-ject \textsl{Uncertainties – Teaching AI its Limits} (BWST\_IF2020-010). IP and DS are supported by the U.S. Department of Energy (DOE), Office of Science grant DOE-SC0010008 and BN is supported by the DOE under contract DE-AC02-05CH11231.

\appendix
\section{Architecture and training}
\label{sec:arch_training}
Here we briefly describe the architecture and training procedure used for \cf~(see~\cite{Krause:2021ilc,Krause:2021wez} for more details). There are some differences compared to the implementation in the original \cf~papers~\cite{Krause:2021ilc,Krause:2021wez}, but most of main algorithm remains the same. 

Both Flow-I and Flow-II are Masked Autoregressive Flows (MAFs)~\cite{papamakarios2017masked} with compositions of Rational Quadratic Splines (RQS)~\cite{durkan2019neural} as transformations. The RQS transformations are parameterized using neural networks known as MADE blocks~\cite{germain2015made}. The details of the architecture of Flow-I and Flow-II are summarized in Table~\ref{tab:flow_architecture}.

\begin{table*}
\begin{center}
\begin{tabular}{|c|c|c|c|c|c|c|c|c|}
\hline
{\multirow{2}{*}{}}&dim of & number of & \multicolumn{3}{c|}{layer sizes} & number of & RQS\\[-0.2ex]
Model &base distribution & MADE blocks & input & hidden & output & RQS bins & tail bound \\
\hline
\hline
Flow-I & 3 & 6 & 64  & $2\times 64$ & 69 & 8 & 14\\
Flow-II & 504 &8 & 378  & $1\times 378$ & 11592 &  8 & 14\\
\hline
\end{tabular}
\caption{Summary of architecture of the Flow-I and Flow-II. For the hidden layer sizes, the first number is the number of hidden layers in each MADE block and the second number is the number of nodes in each hidden layer (e.g., $2\times 64$ refers to 2 hidden layers per MADE block with 64 nodes per hidden layer).}
\label{tab:flow_architecture}
\end{center}
\end{table*}

The incident energy of the incoming photon is preprocessed as 
\begin{align}
E_\text{inc} \to \log_{10} (E_\text{inc}/10 \ \text{GeV})\,.
\end{align}
The inputs to the flows are preprocessed as follows: 

\begin{itemize}
    \item Flow-I: \(E_i \to 2\left(\log_{10}(E_i+1 \ \text{keV})-1\right)\)
    \item Flow-II: 
    \subitem{\(E_i \to \log_{10}(E_i+1 \ \text{keV})-2\),}
    \subitem{\(\mathcal{I}_{ia} \to u_{\text{logit},ia}(\mathcal{I}_{ia}/E_i)\),}
    \subitem{where $u_{\text{logit},ia} = \log\frac{\tilde u_{ia}}{1-\tilde u_{ia}}$,}
    \subitem{$\tilde u_{ia} = \alpha + (1 - 2\alpha)u_{ia}$ and $\alpha = 10^{-6}$
.}
\end{itemize}
The index $i$ denotes the layer number, while the index $a$ specifies the voxel within the given layer. In the original \cf, a different preprocessing was used for the layer energies $E_i$ in Flow-I where $E_i$ were transformed to unit-space (see \cite{Krause:2021ilc}). 
 
As in Ref.~\cite{Krause:2021ilc,Krause:2021wez}, uniform noise in the range [0,1] keV was applied to the voxel energies during training and evaluation. The addition of noise was found to prevent the flow from fitting unimportant features. The training of both flows in this work is optimized using independent \textsc{Adam} optimizers~\cite{kingma2014adam}. Flow-I was trained by minimizing $-\log p(E_0, E_1, E_2|E_\text{inc})$ for 75 epochs with a batch size of 200. Flow-II was trained by minimizing $-\log p\left(\hat{\mathcal{I}}| E_0, E_1, E_2, E_\text{inc}\right)$ for 100 epochs with a batch size of 200. The initial learning of $1\times10^{-4}$ was chosen for the two flows and a multi-step learning schedule was used when training the flows which halves the learning rate after each selected epoch milestone during the training.
\bibliographystyle{JHEP}
\bibliography{HEPML,literature}
\end{document}